\documentclass[aps,prb,preprint,superscriptaddress]{revtex4}

\usepackage[T1]{fontenc}
\usepackage{textcomp}
\usepackage{dcolumn}
\usepackage{bm}
\usepackage{color}
\usepackage{graphicx,subfigure,color}
\usepackage[fleqn]{amsmath}
\usepackage{times}

\begin{document}
\title{Model of Third Harmonic Generation and Electric Field Induced Optical Second Harmonic  using the SBHM}

\author{Adalberto Alejo-Molina}
\affiliation{Direcci\'{o}n de C\'{a}tedras CONACyT, comisionado a: Centro de Investigaci\'{o}n en Ingenier\'{i}a y Ciencias Aplicadas, UAEM Cuernavaca, Mor. 62160, Mexico}

\author{Kurt Hingerl}
\email[email: ]{kurt.hingerl@jku.at}
\affiliation{Center for surface- and nanoanalytics, Johannes Kepler University, Altenbergerstr. 69, 4040 Linz, Austria}

\author{Hendradi Hardhienata}
\affiliation{Center for surface- and nanoanalytics, Johannes Kepler University, Altenbergerstr. 69, 4040 Linz, Austria}

\begin{abstract}
We report for the first time a comprehensive study of the fourth rank tensor describing third harmonic generation (THG) and electric field induced second harmonic  (EFISH) in centrosymmetric material from two different viewpoints: Group Theory (GT) and the Simplified Bond Hyperpolarizability Model (SBHM). We show that the fourth rank tensor related to THG and direct current (DC) EFISH can be reduced to two independent elements whereas SBHM always gives only one, reproducing perfectly well  EFISH experimental results in Metal Oxyde Semiconductor (MOS). We argue that it is possible to reduce the fourth rank tensor describing EFISH to a third rank tensor and further deliver a classical explanation of EFISH regarding symmetry breaking where the  term containing $r^3$ in the potential immediately leads to second harmonic generation (SHG).
\end{abstract}

\maketitle
\date{\today}

\section{Introduction}
Nonlinear optics has become an intensive research area in physics after the discovery of the laser by Townes \textit{et al.} in 1955 [1] and theoretical study of harmonic generation by several seminal papers of Bloembergen \textit{et al.} [2, 3] in 1962. One of the most intensely studied nonlinear optic phenomena is the generation of higher harmonics with frequency $2\omega$ known as second harmonic generation (SHG) first investigated by Franken \textit{et al.} in 1961 [4]. The discovery of third harmonic generation soon followed by Maker and Terhune [5] and subsequently by Ward [6].                     

Bloembergen and coworkers reported a study on the phenomena of electric field induced second harmonic (EFISH) generation [7] and afterwards several efforts, both theoretically [8] and also experimentally [9-11] have been performed to understand the physics of EFISH. Some interesting results include the experimental determination of the hyperpolarizabilities for glass [11] and the separation of the bulk and surface nonlinear contributions [9]. EFISH is generated by a direct current (DC) static field that produces a potential gradient which breaks the initial symmetry of the inversion symmetric material. Therefore the nonlinear polarization due to EFISH is characterized by a fourth rank susceptibility tensor $\overleftrightarrow{\chi}^{(3)}$. 

To the best of our knowledge EFISH has never been analyzed from the view point of the Simplified Bond Hyperpolarizability Model (SBHM) and Group Theory (GT) altogether,  even though the model has been able to successfully predict SHG experimental results of Si surfaces/interfaces [12-14]. The SBHM advantage is that the model can describe SHG generated by a Silicon (Si) surface with only two fitting parameter [13] which is rather different from the standard tensorial approach [15]. In our recent work [16] we have also demonstrated how SBHM and GT are related for several crystal orientations such as Si(111), Si(001), and Si(011) and here we expand SBHM to include THG and EFISH.

In the merit of this objective, this work is organized as follows: In Section II  we discuss group theory results concerning the fourth rank tensor that describes third order nonlinear interactions particularly THG and EFISH. In Section III we derive the corresponding tensor using the SBHM and compare it with GT.  We show in Section IV that due to symmetry breaking in Si(001) and Si(111) facets, the resulting group of symmetry can be explained and interpreted by an effective third rank tensor.  In Section V, a SBHM simulation is applied to fit the MOS experiment in Ref. [10].   Additionally, a classical, \textit{ab initio} EFISH model is presented in Section VI. Finally, the conclusions of this work are briefly stated.

 \section{Group Theory Approach}

Contrary to SBHM, which is basically a phenomenological model to calculate the nonlinear response of a medium, group theory is a mathematical logical construction that can be applied to investigate the symmetry operations allowed by a crystal. Because the physics must be the same upon coordinate transformation, the crystal properties, including the (long wavelength) optical ones, must remain unchanged under symmetry operations. Due to specific crystal orientation, only certain rotations and mirror planes exist. It is well known that rotations and reflections can be represented mathematically by matrices. The set of matrices, which contains all the rotations and mirror planes allowed for a particular crystal, is called a point group.

In general, a third-order nonlinear polarization has the form [17]:
\begin{eqnarray}P_{i}^{(3)}(\omega_{q})=\sum \limits_{jkl}\sum \limits_{mnp}\chi_{ijkl}^{(3)}(\omega_{q},\omega_{m},\omega_{n},\omega_{p})E_{j}(\omega_m)E_{k}(\omega_n)E_{l}(\omega_p)\end{eqnarray}
where the subindex in the frequencies denote that they can be different. According to group theory, the fourth rank tensor $\chi^{(3)}_{ijkl}$ for  bulk silicon  in Eq. (1) belongs to a well known point group symmetry labelled $O_h$. The explicit fourth rank tensor for this group of symmetry is similar to the tensor given in Refs. [18, 19]:
\begin{eqnarray}
\chi_{ijkl}^{(3)}=\left(\begin{array}{ccc}
\left(\begin{array}{ccc}
s_{3333} & 0 & 0\\
0 & s_{3322} & 0\\
0 & 0 & s_{3322}\end{array}\right) & \left(\begin{array}{ccc}
0 & s_{3232} & 0\\
s_{3223} & 0 & 0\\
0 & 0 & 0\end{array}\right) & \left(\begin{array}{ccc}
0 & 0 & s_{3232}\\
0 & 0 & 0\\
s_{3223} & 0 & 0\end{array}\right)\\
\left(\begin{array}{ccc}
0 & s_{3223} & 0\\
s_{3232} & 0 & 0\\
0 & 0 & 0\end{array}\right) & \left(\begin{array}{ccc}
s_{3322} & 0 & 0\\
0 & s_{3333} & 0\\
0 & 0 & s_{3322}\end{array}\right) & \left(\begin{array}{ccc}
0 & 0 & 0\\
0 & 0 & s_{3232}\\
0 & s_{3223} & 0\end{array}\right)\\
\left(\begin{array}{ccc}
0 & 0 & s_{3223}\\
0 & 0 & 0\\
s_{3232} & 0 & 0\end{array}\right) & \left(\begin{array}{ccc}
0 & 0 & 0\\
0 & 0 & s_{3223}\\
0 & s_{3232} & 0\end{array}\right) & \left(\begin{array}{ccc}
s_{3322} & 0 & 0\\
0 & s_{3322} & 0\\
0 & 0 & s_{3333}\end{array}\right)\end{array}\right)\end{eqnarray}
Therefore, a general fourth rank tensor has in total $81$ elements. The four rank tensor in Eq. (2), consists of a $3 \times  3$ matrix, where each component of the matrix consist of $3 \times 3$ matrix elements. As will be shown later, not all components are independent [18, 19]. For clarity, the notation used for representing the fourth rank tensor is explained here. The first index  ``$i$'' in $\chi_{ijkl}$  corresponds to the rows and the second index ``$j$'' to the columns in the main matrix (the external one). It follows then that all the elements in the first row and first column of the external $3 \times  3$ matrix  have $s_{11kl}$   indices, whereas for the second row and third column of the external matrix it will be $s_{23kl}$  and so on. In the same way the indices ``$k$'' and ``$l$'' will correspond to the usual way of  labeling a $3 \times 3$ matrix, namely the rows and columns in the inner $3\times3$ matrix respectively.

\textbf{As proven in Refs. [18-21] for three possible different frequencies, the fourth rank tensor for $O_h$ has at most 4 independent elements}  but in special physical cases could be reduced to 3 [18]. However, for a rotating sample, the tensor needs to be correctly rotated  using the following procedure [18]:
\begin{eqnarray}
s'_{ijkl}\left(\phi\right)=R_{im}\left(\phi\right)R_{jn}\left(\phi\right)R_{ko}\left(\phi\right)R_{lp}\left(\phi\right)s_{mnop}\end{eqnarray}
where  $R_{ab}(\phi)$ is the rotation matrix. One then obtains for a rotation $\phi$ along the $z$-axis the following tensor in its most general form:
\begin{eqnarray}\resizebox{1.0 \textwidth}{!} {$
\left(\begin{array}{ccc}
\left(\begin{array}{ccc}
\frac{1}{4}a_{1}-\frac{1}{4}a_{2}\cos(4\phi) & -\frac{1}{4}a_{2}\sin(4\phi) & 0\\
-\frac{1}{4}a_{2}\sin(4\phi) & \frac{1}{2}a_{3}+\frac{1}{2}a_{2}\cos^{2}(2\phi) & 0\\
0 & 0 & s_{3322}\end{array}\right) & \left(\begin{array}{ccc}
-\frac{1}{4}a_{2}\sin(4\phi) & -\frac{1}{2}a_{4}+\frac{1}{2}a_{2}\cos^{2}(2\phi) & 0\\
\frac{1}{2}a_{5}+\frac{1}{2}a_{2}\cos^{2}(2\phi) & \frac{1}{4}a_{2}\sin(4\phi) & 0\\
0 & 0 & 0\end{array}\right) & \left(\begin{array}{ccc}
\text{0} & 0 & s_{3232}\\
0 & 0 & 0\\
s_{3223} & 0 & 0\end{array}\right)\\
\left(\begin{array}{ccc}
-\frac{1}{4}a_{2}\sin(4\phi) & \frac{1}{2}a_{5}+\frac{1}{2}a_{2}\cos^{2}(2\phi) & 0\\
-\frac{1}{2}a_{4}+\frac{1}{2}a_{2}\cos^{2}(2\phi) & \frac{1}{4}a_{2}\sin(4\phi) & 0\\
0 & 0 & 0\end{array}\right) & \left(\begin{array}{ccc}
-\frac{1}{2}a_{3}+\frac{1}{2}a_{2}\cos^{2}(2\phi) & \frac{1}{4}a_{2}\sin(4\phi) & 0\\
\frac{1}{4}a_{2}\sin(4\phi) & \frac{1}{4}a_{1}-\frac{1}{4}a_{2}\cos(4\phi) & 0\\
0 & 0 & s_{3322}\end{array}\right) & \left(\begin{array}{ccc}
\text{0} & 0 & 0\\
0 & 0 & s_{3232}\\
0 & s_{3223} & 0\end{array}\right)\\
\left(\begin{array}{ccc}
\text{0} & 0 & s_{3223}\\
0 & 0 & \text{0}\\
s_{3232} & 0 & 0\end{array}\right) & \left(\begin{array}{ccc}
\text{0} & 0 & 0\\
0 & 0 & s_{3223}\\
0 & s_{3232} & 0\end{array}\right) & \left(\begin{array}{ccc}
s_{3322} & 0 & 0\\
0 & s_{3322} & \text{0}\\
0 & 0 & s_{3333}\end{array}\right)\end{array}\right)$}\end{eqnarray}

where
\begin{eqnarray*}
\begin{array}{c}
a_{1}=s_{3223}+s_{3232}+s_{3322}+3s_{3333}\\
a_{2}=s_{3223}+s_{3232}+s_{3322}-s_{3333}\\
a_{3}=s_{3223}+s_{3232}-s_{3322}-s_{3333}\\
a_{4}=s_{3223}-s_{3232}+s_{3322}-s_{3333}\\
a_{5}=s_{3223}-s_{3232}-s_{3322}+s_{3333}\end{array}\end{eqnarray*}

\subsection{Third Harmonic Generation}

For the case of third harmonic generation (THG) as long as only a monochromatic frequency is used the nonlinear polarization in Eq. (1) takes a simpler form:
\begin{eqnarray}P_{i}^{(3)}(3\omega)=\sum \limits_{jkl}\chi_{ijkl}^{(3)}(3\omega,\omega,\omega,\omega)E_{j}(\omega)E_{k}(\omega)E_{l}(\omega)\end{eqnarray}
where the fields labeled with ``$j$", ``$k$'' and ``$l$'' are now undistinguishable. Because the fields are the same they cannot be distinguished by the experimenter and the tensor can be "symmetrized". Therefore it is allowed to perform intrinsic permutation of the last three indices [17]:
\begin{eqnarray}\chi_{ijkl}=\chi_{ijlk}=\chi_{iklj}=\chi_{ikjl}=\chi_{ilkj}=\chi_{iljk} \end{eqnarray} When this permutation is performed on Eq. (2) we obtain the following tensor
\begin{eqnarray}\resizebox{.7 \textwidth}{!} {$
\chi_{ijkl}^{(3)}(3\omega)=\left(\begin{array}{c}
\begin{array}{ccc}
\left(\begin{array}{ccc}
s_{3333} & 0 & 0\\
0 & c_{1} & 0\\
0 & 0 & c_{1}\end{array}\right) & \left(\begin{array}{ccc}
0 & c_{1} & 0\\
c_{1} & 0 & 0\\
0 & 0 & 0\end{array}\right) & \left(\begin{array}{ccc}
0 & 0 & c_{1}\\
0 & 0 & 0\\
c_{1} & 0 & 0\end{array}\right)\\
\left(\begin{array}{ccc}
0 & c_{1} & 0\\
c_{1} & 0 & 0\\
0 & 0 & c_{1}\end{array}\right) & \left(\begin{array}{ccc}
c_{1} & 0 & 0\\
0 & s_{3333} & 0\\
0 & 0 & c_{1}\end{array}\right) & \left(\begin{array}{ccc}
0 & 0 & 0\\
0 & 0 & c_{1}\\
0 & c_{1} & 0\end{array}\right)\\
\left(\begin{array}{ccc}
0 & 0 & c_{1}\\
0 & 0 & 0\\
c_{1} & 0 & 0\end{array}\right) & \left(\begin{array}{ccc}
0 & 0 & 0\\
0 & 0 & c_{1}\\
0 & c_{1} & 0\end{array}\right) & \left(\begin{array}{ccc}
c_{1} & 0 & 0\\
0 & c_{1} & 0\\
0 & 0 & s_{3333}\end{array}\right)\end{array}\end{array}\right)$}\end{eqnarray}
where $c_{1}=\frac{1}{3}\left(s_{3223}+s_{3232}+s_{3322}\right)$ or $s_{3223}=s_{3232}=s_{3322}$. \textbf{Therefore for THG the general tensor that previously consists of 4 independent components is now reduced to two independent components.}

\subsection{EFISH}
We now consider the case of a static DC field along the $z$ axis, and assume one monochromatic incident field. The nonlinear polarization in Eq. (1) for this particular situation can be stated as:
\begin{eqnarray}P_{i}^{(3)}(2\omega)=\sum \limits_{jkl}\chi_{ijkl}^{(3)}(2\omega,\omega,\omega,0)E_{j}(\omega)E_{k}(\omega)E_{l}(0)\end{eqnarray}
where in this case ``$l$'' has only the $z$ component different from zero. Because of a monochromatic incident field the instrinsic permutation can again be applied for the two middle indices ``$j$'' and ``$k$'' resulting in the following tensor:
\begin{eqnarray}\resizebox{.7 \textwidth}{!} {$
\chi_{ijkl-EFISH}^{(3)}=\left(\begin{array}{c}
\begin{array}{ccc}
\left(\begin{array}{ccc}
s_{3333} & \text{0} & 0\\
0 & c_{1} & 0\\
0 & 0 & c_{1}\end{array}\right) & \left(\begin{array}{ccc}
0 & c_{1} & 0\\
s_{3223} & 0 & 0\\
0 & 0 & 0\end{array}\right) & \left(\begin{array}{ccc}
0 & 0 & c_{1}\\
0 & 0 & 0\\
s_{3223} & 0 & 0\end{array}\right)\\
\left(\begin{array}{ccc}
0 & s_{3223} & 0\\
c_{1} & 0 & 0\\
0 & 0 & 0\end{array}\right) & \left(\begin{array}{ccc}
c_{1} & 0 & 0\\
0 & s_{3333} & 0\\
0 & 0 & c_{1}\end{array}\right) & \left(\begin{array}{ccc}
0 & 0 & 0\\
0 & 0 & c_{1}\\
0 & s_{3223} & 0\end{array}\right)\\
\left(\begin{array}{ccc}
0 & 0 & s_{3223}\\
0 & 0 & 0\\
c_{1} & 0 & 0\end{array}\right) & \left(\begin{array}{ccc}
0 & 0 & 0\\
0 & 0 & s_{3223}\\
0 & c_{1} & 0\end{array}\right) & \left(\begin{array}{ccc}
c_{1} & 0 & 0\\
0 & c_{1} & 0\\
0 & 0 & s_{3333}\end{array}\right)\end{array}\end{array}\right)$}\end{eqnarray}

where $c_{1}=\frac{1}{2}\left(s_{3232}+s_{3322}\right)$. \textbf{Thus the fourth rank tensor describing EFISH consists of 3 independent elements}. One is tempted at this point to apply Kleinman symmetry for  the tensor in Eq. (9) where for the nonresonant condition the nonlinear susceptibility is independent of the frequency permutation and one can freely interchange all the tensor indices [17]. Therefore we have $s_{3232}=s_{3223}$  and we now have only 2 independent components.  There are however reports that such a treatment for EFISH cannot always be performed [22-24]. 

\section{SBHM analysis of THG and EFISH}

SBHM is a classical phenomenological model first constructed by Aspnes and coworkers that can be used to calculate the nonlinear response of anharmonic dipoles where, in its most simplified version, the electrons are assumed to oscillate only in the direction of the bonds [13].  In the case of nonlinear generation inside the bulk, the polarizabilities and hyperpolarizabilities for a monochromatic input field have the same value along all bond directions but to  surface bonds, a different value is assigned due to symmetry breaking in one particular direction. 

However, symmetry breaking $\textit{in the bulk}$ can also occur if there is a static electric field applied in some specific direction. Let us consider the case where a monochromatic electric field is incoming on a Si surface that is rotated along the $z$ axis. Here, without a DC field, there are no SHG contributions from two photon absorption or dipole contribution inside the bulk because the Si atomic cell is centrosymmetric ($2 \times 4$ opposing bonds) and therefore demands $\chi^{(2)}$ to be zero. However, there are several other mechanism that can generate second harmonic signals inside the bulk [14], one of them is EFISH which breaks this centrosymmetricity and can be evaluated using SBHM. 

In fact, Peng and Aspnes [25] already describe THG using SBHM. They focus their analysis on the far field radiation, calculating and discussing in detail the resulting electric field while outlining the steps followed to get there through SBHM and therefore omitting the information richness that can be obtained from the tensorial description of the problem. They also showed that even if a transversal contribution from the bonds is considered, this contribution can be expressed as a constant multiplying the electric field plus exactly the same functional form of the contribution along the bonds. Thus, for this reason, our analysis can be done only taking account of the contribution along the bonds.
From the viewpoint of the SBHM the fourth rank tensor $\overleftrightarrow{\chi}^{(3)}$  takes the form:
\begin{eqnarray}
\overleftrightarrow{\chi}^{(3)}=\frac{1}{V}\sum \limits_{j=1}\alpha_{3}(\omega,\omega,\omega)\left(\mathbf{R}^{(z)}\cdot\hat{b}_{j}\right)\otimes \left(\mathbf{R}^{(z)}\cdot\hat{b}_{j}\right)\otimes\left(\mathbf{R}^{(z)}\cdot\hat{b}_{j}\right) \otimes\left(\mathbf{R}^{(z)}\cdot\hat{b}_{j}\right)\end{eqnarray}	 	
here $V$ is the volume, $\alpha_{3}$ are the third order hyperpolarizabilities, $\hat{b}_{j}$ are the unit vectors in the direction of the atomic bonds, and  $\mathbf{R}^{(z)}$ is the rotation matrix about the $z$ axis. The summation is performed over all eight bonds which are needed to represent the correct response of the conventional cell. If there exist an applied external DC field along a particular direction e.g. in a  metal oxide semiconductor (MOS) structure [10] then the nonlinear polarization in Eq. (1) can be written as:
\begin{eqnarray}\mathbf{P}=\overleftrightarrow{\chi}^{(3)} \cdot \cdot \cdot \mathbf{E}(\omega) \otimes \mathbf{E}(\omega) \otimes \mathbf{E}(0)\end{eqnarray}
where there are two electric fields oscillating at frequency $\omega$ and one static or DC field. 

Before going further, we are going to show that SBHM always generate a tensor with Kleinman symmetry.  Indeed it can be shown e.g. using brute force by applying the most general bond vector components that the final tensor is always symmetric. This is due to the following fact: the tensor is generated by direct product of the same bonds $\hat{b}_j \otimes \hat{b}_j \otimes \hat{b}_j \otimes \hat{b}_j$ and the hyperpolarizabilities in SBHM  are just constants. It was already proven by McGilp that SBHM has Kleinman symmetry in the particular case of SHG for surfaces [26]. Here, our line of arguing is general and valid for any harmonic generation  driven with a single driving frequency (SHG, THG, FHG, ...) and applies for both bulk or surface. Without loss of generality, we are going to take  Eq. (10) only with the outer product of the vectors, therefore:
\begin{eqnarray}
\overleftrightarrow{\chi}^{(3)}=\hat{b}_{j}\otimes\hat{b}_{j}\otimes\hat{b}_{j}\otimes\hat{b}_{j}\end{eqnarray}
which is a fourth rank tensor and in terms of its components can be expressed as:
\begin{eqnarray}
\overleftrightarrow{\chi}_{qrst}^{(3)}=b_{q}b_{r}b_{s}b_{t}\end{eqnarray}
When the indices $q,r,s,$ and $t$ in the fourth rank tensor $\chi^{(3)}$ in Eq. (13) are permuted, the bond components $b_{q},b_{r},b_{s},$ and $b_{t}$ are also permuted in the same way but because the components are just scalars the product result will always be the same. Therefore any permutation of the subindices generates exactly the same tensor $\chi^{(3)}$ and this is Kleinman symmetry. In addition, the tensorial product in Eq. (12) was derived from the one dimensional classical equation of motion in SBHM [13] which assumes motion only along the bonds or it was taken for granted that the electronic potential is very much aligned along the bonds and even if we wish to take account of the transversal contribution, as was mentioned before this is reduced to a constant multiplying the electric field plus the very same longitudinal contribution along the bonds [25].
We now proceed further to discuss the particular case of silicon. The diamond bond direction inside the Si bulk consists of 8 vectors and the fixed system of reference can be seen in Fig. 1. We choose for the orientation the "standard" conventional diamond cell because the corresponding material tensor for different crystallographic cells in Refs. [18,19] is defined in this way and their representation changes when the sample is rotated. Therefore we apply the following bond definition [16]: 
\begin{eqnarray*}
\hat{b}_{1}=\left(\begin{array}{c}
-\frac{1}{\sqrt{2}}\sin\frac{\beta}{2}\\
-\frac{1}{\sqrt{2}}\sin\frac{\beta}{2}\\
-\cos\frac{\beta}{2}\end{array}\right)\;\quad \hat{b}_{2}=\left(\begin{array}{c}
\frac{1}{\sqrt{2}}\sin\frac{\beta}{2}\\
\frac{1}{\sqrt{2}}\sin\frac{\beta}{2}\\
-\cos\frac{\beta}{2}\end{array}\right)\end{eqnarray*}
\begin{eqnarray}
\hat{b}_{3}=\left(\begin{array}{c}
-\frac{1}{\sqrt{2}}\sin\frac{\beta}{2}\\
\frac{1}{\sqrt{2}}\sin\frac{\beta}{2}\\
-\cos\frac{\beta}{2}\end{array}\right)\; \hat{b}_{4}=\left(\begin{array}{c}
\frac{1}{\sqrt{2}}\sin\frac{\beta}{2}\\
-\frac{1}{\sqrt{2}}\sin\frac{\beta}{2}\\
\cos\frac{\beta}{2}\end{array}\right)\end{eqnarray}
with opposing bonds: 
\begin{eqnarray}
\hat{b}_5=-\hat{b}_1\quad\hat{b}_6=-\hat{b}_2\quad\hat{b}_7=-\hat{b}_3\quad\hat{b}_8=-\hat{b}_4 \end{eqnarray}

  \begin{figure} [htbp]
\includegraphics[scale=0.9]{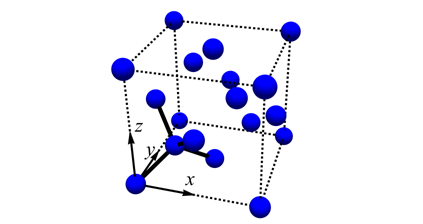}
  \caption{Conventional Cell of Silicon}.
  \label{fig:figure3}
\end{figure}

Using the SBHM method in obtaining the tensor (Eq. 10) and following the index convention, the fourth rank tensor $\overleftrightarrow{\chi}^{(3)}$   takes the form:
\begin{eqnarray}\resizebox{.9 \textwidth}{!} {$
\overleftrightarrow{\chi}^{(3)}=\frac{1}{V}\left(\begin{array}{ccc}
\left(\begin{array}{ccc}
\frac{4}{9}\alpha_{3}\left[3-\cos\left(4\phi\right)\right] & -\frac{4}{9}\alpha_{3}\sin\left(4\phi\right) & 0\\
-\frac{4}{9}\alpha_{3}\sin\left(4\phi\right) & \frac{8}{9}\alpha_{3}\cos^{2}\left(2\phi\right) & 0\\
0 & 0 & \frac{8}{9}\alpha_3\end{array}\right) & \left(\begin{array}{ccc}
-\frac{4}{9}\alpha_{3}\sin\left(4\phi\right) & \frac{8}{9}\alpha_{3}\cos^{2}\left(2\phi\right) & 0\\
\frac{8}{9}\alpha_{3}\cos^{2}\left(2\phi\right) &\frac{4}{9}\alpha_{3}\sin\left(4\phi\right) & 0\\
0 & 0 & 0\end{array}\right) & \left(\begin{array}{ccc}
0 & 0 & \frac{8}{9}\alpha_{3}\\
0 & 0 & 0\\
\frac{8}{9}\alpha_{3} & 0 & 0\end{array}\right)\\
\left(\begin{array}{ccc}
-\frac{4}{9}\alpha_{3}\sin\left(4\phi\right) & \frac{8}{9}\alpha_{3}\cos^{2}\left(2\phi\right) & 0\\
\frac{8}{9}\alpha_{3}\cos^{2}\left(2\phi\right) & \frac{4}{9}\alpha_{3}\sin\left(4\phi\right) & 0\\
0 & 0 & 0\\
\end{array}\right) & \left(\begin{array}{ccc}
\frac{8}{9}\alpha_{3}\cos^{2}\left(2\phi\right) & \frac{4}{9}\alpha_{3}\sin\left(4\phi\right) & 0\\
\frac{4}{9}\alpha_{3}\sin\left(4\phi\right) & \frac{4}{9}\alpha_{3}\left[3-\cos\left(4\phi\right)\right] & 0\\
0 & 0 & \frac{8}{9}\alpha_{3}\end{array}\right) & \left(\begin{array}{ccc}
0 & 0 & 0\\
0 & 0 & \frac{8}{9}\alpha_{3}\\
0 & \frac{8}{9}\alpha_{3} & 0\end{array}\right)\\
\left(\begin{array}{ccc}
0 & 0 & \frac{8}{9}\alpha_{3}\\
0 & 0 & 0\\
\frac{8}{9}\alpha_{3} & 0 & 0\end{array}\right) & \left(\begin{array}{ccc}
0 & 0 & 0\\
0 & 0 & \frac{8}{9}\alpha_{3}\\
0 & \frac{8}{9}\alpha_{3} & 0\end{array}\right) & \left(\begin{array}{ccc}
\frac{8}{9}\alpha_{3} & 0 & 0\\
0 & \frac{8}{9}\alpha_{3} & 0\\
0 & 0 & \frac{8}{9}\alpha_{3}\end{array}\right)\end{array}\right)$}\end{eqnarray}
where  $\alpha_{3}$ is the third order hyperpolarizability for the bulk and the trigonometric functions of angle $\beta$ were evaluated in order to simplify the expression. Comparison between Eq. (16) and Eq. (4) shows similarities in the tensor components. However, even though this procedure can in principle be done for a general angle $\phi$, the result is very cumbersome. Nevertheless, we can set $\phi =0$ and compare it with Eq. (7):
\begin{eqnarray}\resizebox{.5 \textwidth}{!} {$
\overleftrightarrow{\chi}^{(3)}=\frac{8\alpha_{3}}{9V}\left(\begin{array}{ccc}
\left(\begin{array}{ccc}
1 & 0 & 0\\
0 & 1 & 0\\
0 & 0 & 1\end{array}\right) & \left(\begin{array}{ccc}
0 & 1 & 0\\
1 & 0 & 0\\
0 & 0 & 0\end{array}\right) & \left(\begin{array}{ccc}
0 & 0 & 1\\
0 & 0 & 0\\
1 & 0 & 0\end{array}\right)\\
\left(\begin{array}{ccc}
0 & 1 & 0\\
1 & 0 & 0\\
0 & 0 & 0\end{array}\right) & \left(\begin{array}{ccc}
1 & 0 & 0\\
0 & 1 & 0\\
0 & 0 & 1\end{array}\right) & \left(\begin{array}{ccc}
0 & 0 & 0\\
0 & 0 & 1\\
0 & 1 & 0\end{array}\right)\\
\left(\begin{array}{ccc}
0 & 0 & 1\\
0 & 0 & 0\\
1 & 0 & 0\end{array}\right) & \left(\begin{array}{ccc}
0 & 0 & 0\\
0 & 0 & 1\\
0 & 1 & 0\end{array}\right) & \left(\begin{array}{ccc}
1 & 0 & 0\\
0 & 1 & 0\\
0 & 0 & 1\end{array}\right)\end{array}\right)$}\end{eqnarray}

\textbf{It is straightforward to see from Eq. (17) that the SBHM only demands 1 independent parameter which is the third order  hyperpolarizability $\alpha_3$}. In this way, the undetermined GT constants from the independent elements can be determined in terms of physical values. Because the tensor in Eq. (17) is general, the bond model thus predicts that THG and EFISH can also be described by only one independent parameter.  As we have shown in the previous section,  the group theory fourth rank tensor for THG and EFISH   can be reduced, respectively, to only 2 and 3 independent parameters. Therefore it is very possible that the independent tensor elements obtained by GT can be further reduced.  It is now up to the experiment or an \textit{ab initio} theory, to verify if this is generally true. In fact, we will show in Section V, for the case of EFISH with a DC field along the $z$ axis how SBHM can fit experimental results using only one hyperpolarizability value.

\section{EFISH represented by a third rank tensor}

In this section, we argue further that SHG due to a DC EFISH in a particular axis can also be described using a third rank tensor with an effective susceptibility. The argument is as follows:  It is known that $O_h$ posseses one of the highest possible point group symmetries in GT. This group includes $C_4$, $C_3$, $C_2$, $S_6$ and $S_4$ axis, as well as, $\sigma_h$ and $\sigma_{d}$  mirror planes [18, 21]. Physically, the DC field in the direction normal to the [001] plane or along the $z$ axis, breaks the symmetry in the conventional cell: the electronic distribution is no longer the same in other directions and all the elements of symmetry that transforms in some way the $z$ coordinate are no longer allowed. For this reason, only the axis with symmetry group $C_4$ and two vertical mirror planes remain: the resulting point group is $C_{4v}$. 

This symmetry breaking is very similar to the SHG result obtained for a Si(001) surface eventhough they are not exactly the same. The main difference is that for the Si bulk two neighboring tetrahedral elements are required to model the response of the complete conventional cell whereas only one tetrahedral is required to describe the surface. Therefore, the point group for this surface is $C_{2v}$, as we have discussed in our previous work [16]. Based on this, we can contract the general EFISH tensor in Eq. (2)  with a unitary vector in the $z$ direction due to the DC field alienation along this axis so that we obtain a third rank tensor associated with a Si(001) direction:
\begin{eqnarray}\resizebox{0.45\textwidth}{!} {$
\chi_{ijk,GT-EFISH}^{(2)}=
\left(\begin{array}{c}
\left(\begin{array}{ccc}
0 & 0 & s_{3232}\\
0 & 0 & 0\\
s_{3322} & 0 & 0\end{array}\right)\\
\left(\begin{array}{ccc}
0 & 0 & 0\\
0 & 0 & s_{3232}\\
0 & s_{3322} & 0\end{array}\right)\\
\left(\begin{array}{ccc}
s_{3223} & 0 & 0\\
0 & s_{3223} & 0\\
0 & 0 & s_{3333}\end{array}\right)\end{array}\right)$}\end{eqnarray} 
This third rank tensor can be understood as a $9 \times 3$ matrix divided into three matrices with dimension $3 \times  3$. For a general third rank tensor   (``$i$'', ``$j$'', ``$k$''$= 1, 2, 3$), the first index ``$i$'' corresponds to the rows in the main matrix (the external one). Similarly, the indices ``$j$'' and ``$k$'' will correspond to the usual way of  labeling a $3  \times 3$ matrix, where the indices $"j"$ and $"k"$ are respectively the rows and columns in the inner $3  \times 3$ matrix.  Moreover, for the third rank tensors and the case of SHG, where the two fundamental driving fields are undistinguishable, we can again apply intrinsic permutation and therefore $s_{3322}=s_{3232}$, which was already sugested by Eq. (9). \textbf{Furthermore, assuming symmetry in the diagonal of the contracted matrix representation [19] we have $s_{3223}=s_{3232}$ and again both EFISH and SHG third rank tensor now only requires two independent parameters}. 

We can compare Eq. (18) with the SBHM tensor in Eq. (17) after contracting it with the DC field in $z$ direction:
\begin{eqnarray}\resizebox{0.35 \textwidth}{!} {$
\chi^{(2)}_{ijk,SBHM-EFISH}=\frac{8\alpha_{2eff}}{9V}\left(\begin{array}{c}
\left(\begin{array}{ccc}
0 & 0 & 1\\
0 & 0 & 0\\
1 & 0 & 0\end{array}\right)\\
\left(\begin{array}{ccc}
0 & 0 & 0\\
0 & 0 & 1\\
0 & 1 & 0\end{array}\right)\\
\left(\begin{array}{ccc}
1 & 0 & 0\\
0 & 1 & 0\\
0 & 0 & 1\end{array}\right)\end{array}\right)$}\end{eqnarray}
which gives the same form as Eq. (18).   It has to be emphasized that for the case of EFISH the third order hyperpolarizability now takes the form $\alpha_{2eff}=\alpha_{2}(\omega,\omega,0)$ and can generally differ from $\alpha_{2}(\omega,\omega, \omega)$. However, if the electron  movement can be described by  \textbf{only one} resonator, and this is very probable below the band gap, the static EFISH polarizibility \textbf{has to be connected} to the $\alpha_{2}(\omega,\omega,0)$, just applying an oscillator model. In Section VI we will show how an expression for $\alpha_{2eff}$ can be obtained from symmetry breaking of the atomic potential.

Furthermore, it is also interesting to explore what happens when the DC field is aligned normal to the Si(111) surface because this orientation has been reported experimentally [10] and studied theoretically using SBHM [13, 14]. To achieve this, one needs to transform the general tensor to another reference frame before doing the contraction. This additional step is necessary because the system of reference shown in Fig. 1 is not the same as that used in standard group theory textbooks  [18, 19], where the $z$ axis corresponds to the higher order of rotation of the point group under analysis. To go from the original system of reference, namely in the direction Si(001) to the direction of Si(111), two independent rotations are required. The first is a rotation of $\pi/4$ around the $z$ axis and after that a second rotation of $\beta/2$ about the $x$ axis is performed. One then finally arrives to the Si(111) orientation. The transformation is shown in Fig. 2 but for simplicity it is shown only acting on a tetrahedral element; this tetrahedral element can be seen with the bonds located at the left corner of Fig. 1. 

  \begin{figure} [htbp]
\includegraphics[scale=0.9]{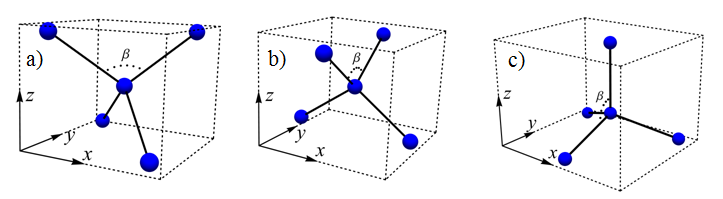}
  \caption{Pictorial description of a tetrahedral element rotation to go from the Si(001) to Si (111) direction. a) SBHM configuration, b) after a $\pi/4$ rotation around $z$ axis and c) a second rotation of $\beta/2$ around $x$ axis}.
  \label{fig:figure3}
\end{figure}

 Matematically, these transformations can be applied to the tensor in Eq. (17) using relation from Eq. (3). First, a rotation is applied about the $z$ axis and evaluated at $\phi = \pi/4$ which we label $\mathbf{R}^{(z)}(\pi/4)$ and after that a second transformation about the $x$ axis for an angle of $\beta/2$ as mentioned before labeled $\mathbf{R}^{(x)}(\beta/2)$. Therefore, after applying these transformations and contracted with a unitary vector in the direction $z$ (in this case the $z$ axis is parallel to the direction of Si(111) as in Fig. 2 c), the resulting effective third rank tensor for the Si(111) orientation is
\begin{eqnarray}
\overleftrightarrow{\chi}^{(2)}_{SBHM-EFISH}=\frac{8\alpha_{2eff}}{27V}\left(\begin{array}{c}
\left(\begin{array}{ccc}
0 & -\sqrt{2} &1\\
-\sqrt{2} & 0 & 0\\
1 & 0 & 0\end{array}\right)\\
\left(\begin{array}{ccc}
-\sqrt{2} & 0 & 0\\
0 & \sqrt{2} & 1\\
0 & 1 & 0\end{array}\right)\\
\left(\begin{array}{ccc}
1 & 0 & 0\\
0 & 1 & 0\\
0 & 0 & 7\end{array}\right)\end{array}\right)\end{eqnarray}
which again shows that in SBHM the third rank tensor describing EFISH requires only one independent parameter which is the second order hyperpolarizability. 

Thus in the same way, when the tensor is transformed from the (001) to (111) direction, the symmetry elements in $O_h$ are reduced only to the $C_{3v}$ axis parallel with $z$. Interestingly from the SBHM viewpoint, the symmetry of a Si(111) orientation is such that even when two tetrahedral elements are used to represent the bulk response, the symmetry remains the same as that of a Si(111) surface (represented by only one tetrahedral element) in a similar manner when we describe the effective suseptibility in GaAs(111) in our previous work [16] but this time due to EFISH. Therefore, the SBHM model only establishes one additional relation between the independent elements in the tensor because the other ones can be derived from the symmetry in the crystal. In addition, modeling EFISH only using one independent element in the tensor is not uncommon at all, Kikuchi and Tada using quantum mechanical perturbative calculations [8] showed that under certain conditions it is possible to use only one independent element in the fourth rank tensor to describe EFISH. In Section V we show that using both the fourth and third rank tensor we can model EFISH perfectly.

\section{SBHM simulation on MOS EFISH EXPERIMENT}

We now show that SBHM predicts the experimental EFISH result for a $pp$ (first letter polarization of the $in$- field, second letter  for -$out$) of  Si(111) as published by Aktsipetrov and coworkers [10] by using only one hyperpolarizability parameter for the fit. This assumption is valid if we consider that the SHG is generated mainly by charge separation in the Si-SiO$_{2}$ interface. Indeed in their transmission experiment, the coherence length between fundamental and SH is long or larger than the size of the depletion region, therefore in the measured signal phase matching as well as the surface effect can be neglected which is in agreement with the arguments of Refs. [18,19]. In addition, internal photoemission between such interface has been studied using time dependent (TD) SHG as an interface dopant probe [27-30]. 

We apply the following bond orientation for the Si(111) bulk [31] which is the same if one starts with the diamond unit cell bond orientation in Eq. (12) and perform the necessary transformation into a Si(111) orientation:
\begin{eqnarray*}
\hat{b}_{1}=-\hat{b}_{5}=\left(\begin{array}{c}
0\\
0\\
1\end{array}\right)\;\quad \hat{b}_{2}=-\hat{b}_{6}=\left(\begin{array}{c}
0\\
-\sin \beta\\
\cos \beta\end{array}\right)\end{eqnarray*}
\begin{eqnarray}
\hat{b}_{3}=-\hat{b}_{7}=\left(\begin{array}{c}
\frac{\sqrt{3}}{2}\sin\beta\\
\frac{1}{2}\sin\beta\\
\cos\beta\end{array}\right)\; \hat{b}_{4}=-\hat{b}_{8}=\left(\begin{array}{c}
-\frac{\sqrt{3}}{2}\sin\beta\\
\frac{1}{2}\sin\beta\\
\cos\beta\end{array}\right)\end{eqnarray}
where $\beta$  is the angle between each bond and is equal to $2\cos^{-1}[1/\sqrt{3}]\approx109.47^{0}$ for silicon. The far field is obtained by [13, 31, 32] 
\begin{eqnarray}\mathbf{E}_{ff}^{2\omega}=\frac{k^{2} e^{ikr}}{r}\left[\hat{\mathbf{I}}-\hat{k}\hat{k}\right]\cdot \sum \limits_{n=1}^{N}\sum \limits_{j=1}\mathbf{p}_{j}^{(2),n}\end{eqnarray}
where $\hat{\mathbf{I}}$  is a $3\times3$ unit tensor, $\mathbf{p}_{j}^{(2),n}$  is the considered SHG polarization source and $\hat{k}$  is the unit vector of the electric field in the direction of the laboratory observer and has the form
\begin{eqnarray}\hat{k}=\cos \theta_{0}\hat{x}+\sin \theta_{0} \hat{z} \end{eqnarray}

Following the setup in [10], we set $\theta_{0}=0$  so the electric field is polarized along the $x$ axis only. Because the intensity is in arbitrary unit we can assign arbitrary constant for the third order hyperpolarizability. The DC bias is varied for every $\pm0.25$V interval  around two reference valleys. We first simulate the SHG azimuthal feature for a zero external DC using Eq. (17). For a zero bias, only the internal field due to 
the formation of a depletion field contributes to SHG. Without this internal field bulk dipole is automatically zero by SBHM simulation because $\hat{b}_{1} = -\hat{b}_{5}$, $\hat{b}_{2} = -\hat{b}_{6}$, $\hat{b}_{3} = -\hat{b}_{7}$, and $\hat{b}_{4} = -\hat{b}_{8}$ and the radiated fields produced by the bond charges are cancelling out.

The simulation result for the $pp$ polarization is given in Fig. 3 and shows the expected 6 fold pattern at an arbitrary unit SHG intensity of $0.4$ (Fig. 3a) where we assume arbitrary constant  internal DC bias (depletion voltage). The physical explanation according to SBHM is that for a normal incidence field the electric field $p$-polarization is parallel to the $x$ axis thus affecting each rotated down-bond $\hat{b}_{2}$, $\hat{b}_{3}$, and $\hat{b}_{4}$  equally producing a maximal bond radiation when each down bond aligns parallel to the $x$ axis (twice when rotated $360^o$) therefore a sixfold SHG pattern occurs. All this occurs naturally using SBHM and one can even look at each single bond contribution before summing them up to get a better physical understanding and found that the up bonds $\hat{b}_1$ and $\hat{b}_5$ do not contribute to any SHG intensity for a normal incidence $pp$ field because the bond direction is perpendicular to the direction of the electric field. This is in our opinion one of the advantages of SBHM because even though limited to the classical view it gives a strong physical picture of how SHG is generated by various sources. The result agrees nicely with the experimental data reported in Fig. 1 of the Ref. [10]. Our simulation further shows that applying an external DC bias does not change the sixfold pattern but increase or decrease the SHG intensity.

 \begin{figure} [htbp]
\includegraphics[width=16cm]{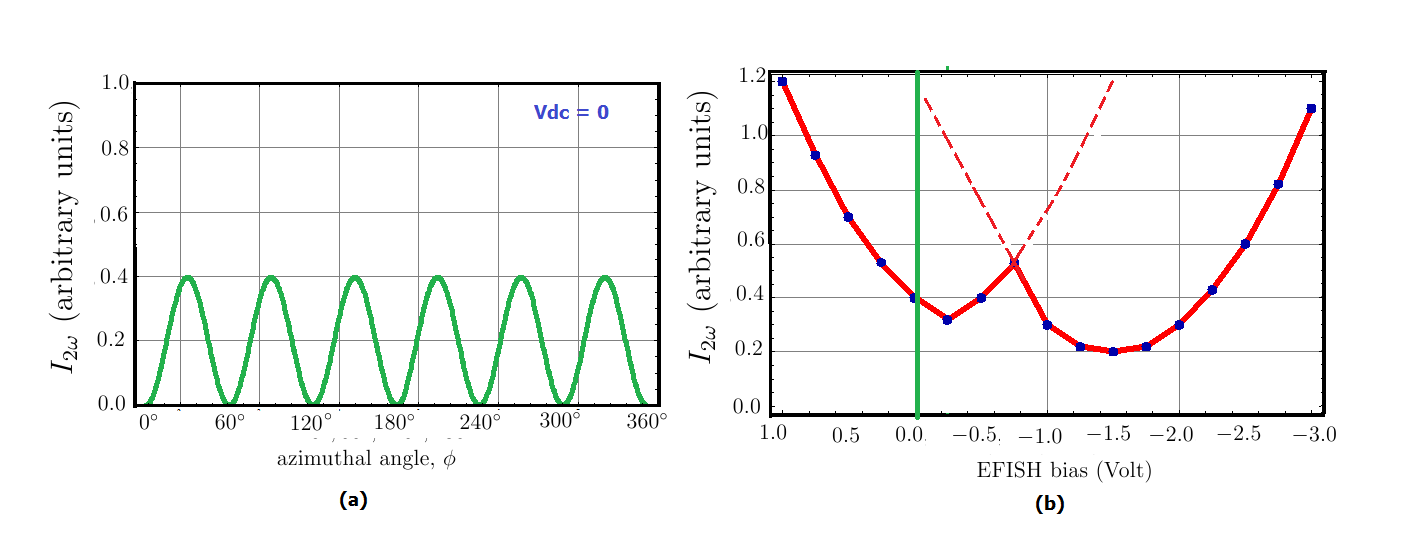}
  \caption{SBHM simulation result for (a) zero external DC bias revealing the sixfold pattern  (b) SHG peak intensity as a function of the DC bias. The first valley is at $V=-0.25$V and the second one is at $V=-1.5$V and is used as the zero volt reference before a forward and backward voltage is applied producing a parabolic curve around the two reference voltages. The simulation agrees very well with the experimental results in Ref. [10]}
  \label{fig:figure3}
\end{figure}

We next vary the DC bias for a given reference voltage. According to Ref. [10] there are two SHG intensity valleys, namely at $V=-0.25$V with a SHG arbitrary unit intensity of around $0.32$ and $V=-1.25$V with a SHG intensity of $0.2$.  Aktsipetrov and coworkers argue that the absence of a single valley is either due to electric charge redistribution inside the SiO$_{2}$ or due to surface states recharging . Whatever is the cause, it somehow shifts the zero bias reference at  $V=-0.25$V to $V=-1.5$V and we adopt this shift in the simulation to produce two symmetric parabolic fits (Fig 3b). Because the SHG intensity is obtained by the square of the field two symmetric parabolic curve around the two valleys are expected and is also produced by the simulation. The simulation result compares well with experiment result in Fig. 2 of Ref. [10]. We will show in Section VI, how this experiment can be fully reproduced using a third rank tensor approach.

\section{Classical Picture of EFISH in Centrosymmetric Crystal}
 
In this section a brief description about a classical process model of EFISH is provided. It is well known that the SHG dipole response in bulk Si(111) without a static electric field is zero if we neglect the biatomic structure of Si and absorption. Such a system can be modeled as having an atomic potential consisting of only even terms [17]. When a static field is directed towards the $z$ axis  it will break the symmetry of the atomic potential. \textbf{From the view of the bond model this static field produces a projection on the three down bonds so that it slightly shifts the equilibrium position of the electron away from the former equilibrium position before the DC field was applied}. Because the lattice position can be assumed to be static in the process this change in equilibrium can be seen as breaking the initial symmetry.
 
 \begin{figure} [htbp]
\begin{center}
\includegraphics{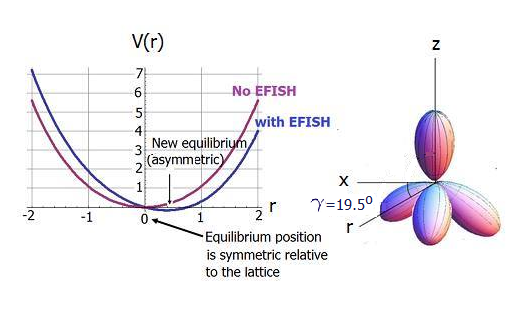}
\end{center}
  \caption{ Classical picture of EFISH in Si(111). Left: A static EFISH will shift the equilibrium position from the symmetric lattice position and the potential now becomes asymmetric relative to the lattice. Right: The projection of the static field on the down bond.}
  \label{fig:figure3}
\end{figure}
 
To the best of our knowledge this picture of the shifting of the potential has never been discussed before. The potential of a centrosymmetric material can be expressed as:
\begin{eqnarray}V(r)=m\omega_{0}r^{2}-mbr^{4}\end{eqnarray}
where here we use $r$ as the radial coordinate in SBHM, which is related to $z$ as $ q E_{dc} r \sin \gamma$, with $\gamma$ as angle between the upward bond directions onto the $z$ axis. Now a static field along the $z$ axis will alter the potential in the form, abbrevating $ E_{DC} = E_{dc} \sin \gamma $
 \begin{eqnarray}V(r)=m\omega_{0}r^{2}-mbr^{4}-q E_{DC} r \end{eqnarray}
here $\omega_{0}$ is the oscillator resonance frequency, $q$ is the electron charge, and $b$ is the parameter that characterizes the third harmonic generation (THG) strength.
 
This field will produce a shift in the electron potential relative to its initial position, for the bond in [111] direction the shift will be larger than for the three backbonds. The force equation is then obtained by differentiating the above equation. The new electron equilibrium positions can be obtained by assuming $b r^4 << \omega_0 r^2 $. The solution is
 
\begin{eqnarray}r=r_{0}-\frac{q E_{DC}}{2\omega_{0}}\end{eqnarray}
 
with $r_{0}$ as initial, symmetric position. We now apply a series expansion for the potential $V(r)$ around the new electron equilibrium position, $r$ and obtain
\begin{eqnarray}\begin{split}
-\frac{bE_{DC}^{4}+4E_{DC}^{2}\omega_{0}^{3}}{16\omega_{0}^{4}}-\frac{bE_{DC}^{3}}{2^{2}\omega_{0}}\left(r-\frac{E_{DC}}{2\omega_{0}}\right)+\left(-\frac{3bE_{DC}^{2}}{2\omega_{0}^{2}}+\omega_{0}\right)\left(r-\frac{E_{DC}}{2\omega_{0}}\right)^{2}\\ -\frac{2bE_{DC}}{\omega_{0}}\left(r-\frac{E_{DC}}{2\omega_{0}}\right)^{3}+O\left[r-\frac{E_{DC}}{2\omega_{0}}\right]^{4}\end{split}\end{eqnarray}
The term in the potential containing $r^3$
\begin{eqnarray}
-\frac{2bE_{DC}}{\omega_{0}}\left(r^{3}-\frac{E_{DC}^{3}}{8\omega_{0}^{3}}+\frac{3E_{DC}^{2}}{4\omega_{0}^{2}}r-\frac{3E_{DC}}{2\omega_{0}}r^{2}\right)\end{eqnarray}
 
when viewed in  the force equation is proportional to $r^2$, thus:
 
\begin{eqnarray}\alpha_{2eff}^{(2)}(2\omega)=\frac{6 b E_{DC}}{\omega_{0}}\end{eqnarray}
It is interesting that based on Eq.  (29) the SHG EFISH susceptibility strength depends on the characterizing parameter $b$ of THG and yields immediately zero if $E_{DC}=0$ as expected. Increasing the DC EFISH voltage will increase the SHG contribution. The linear dependence of $\alpha_{2eff}$ on the DC field is  important because it will result in a parabolic feature in the intensity when the EFISH field is altered. Furthermore SBHM immediately predicts that under normal incidence only the three backbonds contribute to the EFISH signal, despite they experience a rather small shift proportional to $sin(19.5�)$, whereas the top bond along the {111} direction does not contribute, because it is perpendicular to the polarization of the wave. This argument derived from SBHM predicts that for normal incidence on Si(001) EFISH should not occur, because the in-plane symmetry $V(x,y)$ is not broken, despite the electrons move out along the $z$ direction by the static field.
 
It is thus clear that the expression for the SHG intensity which were applied to plot the experiment using the polarization formula in Section V can also be obtained via the third rank tensor in Eq. (20) with an additional rotation freedom about the $z$ axis. Indeed both cases will result in a similar intensity formula yielding a 6 fold dependence for the $pp$ polarization when the applied field is nomal to the surface:
\begin{eqnarray}
I_{pp}\sim E_{DC}^{2}\alpha\sin^{2}{3\phi}\end{eqnarray}
where $\alpha$ can either be represented by the effective first or second order hyperpolarizability. Therefore the experiment in Ref. [10] can be fully refitted to the same accuracy using a third rank tensor.

\section{Conclusion}
We show that the fourth rank tensor related to THG and DC EFISH nonlinear polarization obtained from group theory belongs to the $O_h$ point group  and in its most general form consists of 4 independent parameters which by symmetry arguments can be reduced further to 2   whereas SBHM agrees with all the tensor elements different from zero but only requires one independent element in the fourth rank tensor which is the third order hyperpolarizability. The fourth rank tensor describing DC EFISH in the $z$ axis can be further reduced to a third rank tensor, where GT shows 2 independent parameters and SBHM requires one. We then demonstrate that the DC MOS EFISH experiment in Ref. [10] can indeed be fitted using only one hyperpolarizability value as the fitting parameter. At the end,  we presented an important mechanism, a classical picture of the symmetry breaking and briefly demonstrate how SHG can arise.

\let\cleardoublepage


\clearpage
\begin{thebibliography}{9}

\bibitem{1}
J. P. Gordon, H. J. Zeiger, and C.H. Townes, {}``The maser, new type of microwave amplifier, frequency, standard and spectrometer,{}'' Phys. Rev. \textbf{99}, 1264 -1274 (1955).

\bibitem{2}
N. Bloembergen and P. S. Pershan,{}``Light waves at the boundary of nonlinear media,{}'' Phys. Rev. \textbf{128}, 606 - 622 (1962).

\bibitem{3}
J. A. Armstrong, N. Bloembergen, J. Ducuing, and P. S. Pershan, {}``Interaction between light waves in a nonlinear dielectric,{}'' Phys. Rev. \textbf{127}, 1918 - 1939 (1962).

\bibitem{4}
P. A. Franken, A. E. Hill, C. W. Peters, and G. Weinreich, {}``Generation of optical harmonics,{}''  Phys. Rev. Lett. \textbf{7}, 118 - 119 (1961).

\bibitem{5}
P. D. Maker and R. W. Terhune, {}``Study of Optical Effects Due to an Induced Polarization Third Order in the Electric Field Strength,{}'' Phys. Rev. \textbf{137}, A801-A818 (1965) .

\bibitem{6}
J. F. Ward and G. H. C. New, {}``Optical Third Harmonic Generation in Gases by a Focused Laser Beam{},'' Phys. Rev. \textbf{185},  57-72 (1969).

\bibitem{7}
C. H. Lee, R. K. Chang, and N. Bloembergen, {}``Nonlinear electroreflectance in silicon and silver,{}'' Phys. Rev. Lett. \textbf{18}, 167 -170 (1967).

\bibitem{8}
K. Kikuchi and K. Tada, {}``Theory of electric field-induced optical second harmonic generation in semiconductors,{}'' Opt. Quantum Electron. \textbf{12}, 99-205 (1980).

\bibitem{9}
O. A. Aktsipetrov, A. A. Fedyanin, and A. V. Melnikov, {}``DC electric field induced second-harmonic generation spectroscopy of the Si(001)-SiO$_2$ interface: separation of the bulk and surface non-linear contributions,{}'' Thin Solid Films \textbf{294}, 231-234 (1997).

\bibitem{10}
O. A. Aktsipetrov, A. A. Fedyanin, V. N. Golovkina, and T. V. Murzina, {}``Optical second-harmonic generation induced by a dc electric field at the Si-SiO$_2$ interface,{}'' Opt. Lett. \textbf{19}, 1450-1452 (1994).

\bibitem{11}
C. G. Bethea, {}``Electric field induced second harmonic generation in glass,{}'' App. Optics \textbf{14}, 2435-2437 (1975).

\bibitem{12}
G. L\"{u}pke, D. J. Bottomley, and H. M. van Driel, {}`` Second- and third-harmonic generation from cubic centrosymmetric crystals with vicinal faces: phenomenological theory and experiment,{}'' J. Opt. Soc. Am. B \textbf{11}, 33-44 (1994).


\bibitem{13}
G. D. Powell, J. F. Wang, and D. E. Aspnes, {}``Simplified bondhyperpolarizability model of second harmonic generation,{}''  Phys. Rev. B \textbf{65}, 205320 (2002).

\bibitem{14}
J.-F. T. Wang, G. D. Powell, R. S. Johnson, G. Lucovsky, and D. E. Aspnes, {}``Simplified bond-hyperpolarizability model of second harmonic generation: application to Si-dielectric interfaces,{}'' J. Vac. Sci. Technol. B \textbf{20}, 1699-1705 (2002).

\bibitem{15}
J. E. Sipe, D. J. Moss, and H. M. van Driel, {}``Phenomenological theory of optical second- and third-harmonic generation from cubic centrosymmetric crystals,{}'' Phys. Rev. B \textbf{35}, 1129 - 1141 (1987).

\bibitem{16}
A. Alejo-Molina, H. Hardhienata, and K. Hingerl. {}``Simplified bond-hyperpolarizability model of second harmonic generation, group theory and neumann's principle,{}'' J. Opt. Soc. Am. B \textbf{31}, 526-533 (2014).

\bibitem{17}
R. W. Boyd, \textit{Nonlinear Optics}, 2nd ed. (Academic, 2003).

\bibitem{18}
J. F. Nye, \textit{Physical Properties of Crystals, Their Representations by Tensors and Matrices}, (Clarendon, 1957).

\bibitem{19}
R. C. Powell, \textit{Symmetry, Group Theory, and the Physical Properties of Crystals}, Lecture Notes in Physics (Springer, 2010).

\bibitem{20}
C. C. Shang and H. Hsu, {}``The Spatial Symmetric Forms of Third-Order Nonlinear Susceptibility,{}'' IEEE J. Quantum Electron. \textbf{QE-23}, 177-179 (1987).

\bibitem{21}
D. C. Harris and M. D. Bertolucci, \textit{Symmetry and Spectroscopy-An Introduction to Vibrational and Electronic Spectroscopy} (Dover, 1989).

\bibitem{22}
Christopher A. Dailey, Brian J. Burke, and  Garth J. Simpson, {}``The general failure of Kleinman symmetry in practical nonlinear optical applications,{}'' Chem. Phys. Let. \textbf{390}, 1-3 (2004).

\bibitem{23}
Victor Mizrahi and D. P. Shelton, {} ``Deviations from Kleinman symmetry measured for several simple atoms and molecules,{}'' Phys. Rev. A \textbf{31}, 3145-3154  (1985).

\bibitem{24}
D. P. Shelton and Zhengfang Lui, {} ``Kleinman symmetry deviations for hydrogen,{}'' Phys. Rev. A \textbf{37}, 2231-2233  (1988).

\bibitem{25}
H. S. Peng and D. E. Aspnes, ``Calculation of bulk third-harmonic generation from crystalline Si with the simplified bond hyperpolarizability model,{}'' Phys. Rev. B \textbf{70}, 165312 (2004).
\bibitem{26}
J. F. McGilp, ``Using steps at the Si-SiO$_2$  interface to test simple bond models of the optical second-harmonic response,'' J. Phys. \textbf{19}, 016006 (2007).

\bibitem{27}
J. G. Mihaychuk, N.  Shamir, and H. M. van Driel,  {}``Multiphoton photoemission and electric-field-induced optical second-harmonic generation as probes of charge transfer across the Si/SiO$_2$ interface,{}'' Phys. Rev. B \textbf{59}, 2164-2173 (1999).

\bibitem{28}
Julie L. Fiore, Vasiliy V. Fomenko, Dora Bodlaki, and Eric Borgueta,  {}``Second harmonic generation probing of dopant type and density at the Si/SiO$_2$ interface,{}'' Appl. Phys. Lett. \textbf{98}, 041905 (2011).

\bibitem{29}
H. Park, B. Choi, A. Steigerwald, K. Varga and N. Tolk,  {}``Annealing effect in boron-induced interface charge traps in Si/SiO2 systems,{}'' J. Appl. Phys. \textbf{113}, 023711 (2013).

\bibitem{30}
Yong Q. An, J. Price, Ming Lei and M. C. Downer,  {}``Role of photo-assisted tunneling in time-dependent second-harmonic generation from Si surfaces with ultrathin oxides,{}'' Appl. Phys. Lett. \textbf{102}, 051602 (2013).

\bibitem{31}
H. Hardhienata, A. Prylepa, D. Stifter, and K. Hingerl, {}``Simplified bond-hyperpolarizability model of second-harmonic-generation in Si(111): theory and experiment,{}'' J. Phys. \textbf{423}, 012046 (2013).

\bibitem{32}
J. Kwon, M. C. Downer, and B. S. Mendoza,{}``Second-harmonic and reflectance-anisotropy spectroscopy of vicinal Si(001)/SiO$_2$ interfaces: experiment and simplified microscopic model,{}'' Phys. Rev. B \textbf{73}, 195330 (2006).


\end{thebibliography}
\end{document}